\documentclass[preprint,final,5p,times,twocolumn]{elsarticle}
\usepackage[colorlinks,linkcolor=blue,anchorcolor=blue,citecolor=blue,urlcolor=blue]{hyperref}

\usepackage{natbib,units}

\usepackage{graphicx}
\usepackage{dcolumn}
\usepackage{bm}        
\usepackage{amssymb}   

\usepackage{bm}
\usepackage[usenames ,dvipsnames]{xcolor}
\usepackage{color}
\usepackage{url}
\usepackage{dsfont}        
\usepackage{slashed}
\usepackage{epsfig}
\usepackage{multirow,array}
\usepackage{float}
\usepackage{fancyhdr}
\usepackage{indentfirst}
\usepackage{cancel}
\usepackage{tabularx}
\usepackage{simplewick}
\usepackage{amsmath}
\usepackage[normalem]{ulem}
\usepackage{slashed}
\usepackage{upgreek}
\usepackage{xtab,afterpage,longtable}

\emergencystretch=\maxdimen
\usepackage{dcolumn}
\newcolumntype{x}{D{.}{.}{6.6}}
\newcolumntype{y}{D{.}{.}{4.5}}
\newcolumntype{z}{D{.}{.}{5.7}}
\newcolumntype{f}{D{.}{.}{7.9}}
\newcolumntype{e}{D{.}{.}{5.6}}

\journal{Physics Letters B}

\begin{document}
\begin{frontmatter}

\title{Enhanced Early Galaxy Formation in JWST from Axion Dark Matter?}

\author[UCR]{Simeon Bird}\ead{sbird@ucr.edu}
\author[UCR]{Chia-Feng Chang}\ead{chiafeng.chang@email.ucr.edu}
\author[UCR]{Yanou Cui}\ead{yanou.cui@ucr.edu}
\author[UCR,PMO]{Daneng Yang}\ead{danengy@ucr.edu}

\fntext[alphabetical]{Authors listed in alphabetical order.}

\address[UCR]{Department of Physics and Astronomy, University of California, Riverside, CA 92521, USA}
\address[PMO]{Purple Mountain Observatory, Chinese Academy of Sciences, Nanjing 210033, China}

\begin{abstract}
We demonstrate that enhanced early galaxy formation can generically arise in axion-like particle (ALP) dark matter (DM) models with a delayed onset of axion field oscillation. In these models, the formation of localized massive objects enhances structure formation, potentially addressing the excess recently observed by the James Webb Space Telescope (JWST), while remaining consistent with existing constraints. 
We identify viable parameter space with the ALP mass in the range of $10^{-22}~{\rm eV}<m_a<10^{-19}~\rm eV$. In addition, we show that the ALP parameter regions of interest can lead to intriguing complementary signatures in the small scale structure of DM halos and existing experimental searches for ALPs.
\end{abstract}

\end{frontmatter}

\section{Introduction}
The standard $\Lambda$CDM cosmology makes firm predictions for the abundance of dark matter halos as a function of time. However, recent James Webb Space Telescope (JWST) observations have revealed what may be an unexpectedly large population of luminous galaxies at redshifts $10$ and above~\cite{2023Natur.616..266L,2023MNRAS.519.1201A,2022ApJ...940L..55F,2023ApJS..265....5H,2022arXiv221001777B,2023arXiv230406658H,2006SSRv..123..485G}. In particular, Ref. \cite{2023arXiv230406658H} reported $25$ spectroscopically confirmed galaxies at $z_{\rm spec} = 8.61-13.20$, two of which have $M_{\rm UV}<-19.8~\rm mag$ at $z>11$. 
The reported number exceeds most predictions based on the $\Lambda$CDM cosmology, and may be the harbinger of new fundamental physics and/or lead to a new understanding of structure formation
\cite{2014MNRAS.445.2545D,2019MNRAS.486.2336D,2020MNRAS.496.4574Y,2020MNRAS.499.5702B,2023MNRAS.519.3118W,2023MNRAS.521..497M,2023MNRAS.518.2511L,Boylan-Kolchin:2022kae}.

These surprising JWST results have inspired many dedicated studies
~\cite{2023arXiv230507049S,2023MNRAS.518.2511L,Boylan-Kolchin:2022kae,Chen:2023ugq,2023arXiv230413755M,2022arXiv221010066K}. 
Proposed explanations for the JWST excess include enhanced star formation, accelerated mass assembly, early or clustering dark energies, large scale-dependent non-Gaussianities, cosmic string loops, etc.~\cite{Dekel:2023ddd,Sun:2023ocn,2023MNRAS.521.3201N,Passaglia:2021jla,Biagetti:2022ode,Liu:2022okz,Liu:2022bvr,Hutsi:2022fzw,Batista:2017lwf,2021MNRAS.504..769K,Jiao:2023wcn}.
While it is possible that the JWST excess could be explained within the $\Lambda$CDM framework, it still requires refinements in our understanding of star formation efficiency~\cite{Dekel:2023ddd,Sun:2023ocn,Sun:2023ocn,2022ApJ...939L..31H}. Therefore, proposals based on new physics are strong contenders as alternative explanations, such as a population of highly massive compact objects ~\cite{Zurek:2006sy,Hutsi:2022fzw,Liu:2022okz,Liu:2022bvr}, which is the focus of this work.
The presence of these objects enhances the matter power spectrum (MPS) with a shot-noise-like contribution up to a certain truncation scale~\cite{Carr:2018rid}. 
This shot noise enhancement could then trigger higher star formation efficiency, particularly in the most massive halos and at earlier epochs~\cite{Zurek:2006sy,Dekel:2023ddd,Irsic:2019iff}.
To fully address the excess, however, these objects must be extremely massive and contribute a substantial fraction to the total matter content in the universe ($\Omega_m$) ~\cite{Hutsi:2022fzw,Liu:2022okz,Liu:2022bvr}. 

A well-motivated source of such heavy compact objects is oscillatons arising from theories of axion dark matter (DM) with delayed onset of field oscillations~\cite{Soda:2017dsu,Kitajima:2018zco,Olle:2019kbo,Brax:2020oye}. Parametric resonance in these theories leads to the formation of oscillons~\cite{Urena-Lopez:2001zjo,Marsh:2015xka,Bogolyubsky:1976nx,Copeland:1995fq,Fodor:2006zs,2011PhRvD..84d3531C,Mukaida:2016hwd,Zhang:2018slz,Olle:2020qqy,Olle:2019kbo}, which are localized long-lived objects that may either decay or persist until the present day. In most cases, these oscillons remain localized and become oscillatons~\cite{Visinelli:2021uve,Ikeda:2017qev,Fodor:2009kg} supported by gravity by $z \sim 10$~\cite{Hogan:1988mp,Kolb:1993zz,Kolb:1993hw,Kolb:1994fi,Niemeyer:2019aqm,Barman:2021rdr,Enander:2017ogx}, with masses ranging from $10^{-12}~\rm M_{\odot}$ to $10^4~\rm M_{\odot}$~\cite{Zurek:2006sy,Fairbairn:2017sil}. 
The mass of these oscillatons anti-correlates with the axion mass $m_a$. 
The $m_a$ range favored by the JWST excess is $10^{-19} ~{\rm eV} <m_a<10^{-16}~\rm eV$ in the Standard Misalignment Mechanism (SMM) \cite{Hutsi:2022fzw}, yet further requires a larger star formation efficiency $f_*$ than expected from low redshift astrophysics. However, this mass range is excluded by black hole super-radiance (BHSR)~\cite{Arvanitaki:2014wva, Hutsi:2022fzw,Stott:2018opm,Unal:2020jiy,Kimball2023}. 

In this work, we show that models with delayed axion oscillation can open up a parameter space safe from the BHSR constraint with astrophysically plausible $f_*$. 
In particular, we reveal a large viable parameter region with $10^{-22}~{\rm eV} < m_a < 10^{-19} \rm eV$ that can explain the JWST excess, while being consistent with relevant constraints such as from BHSR, Lyman-$\alpha$ forest, and stellar dynamics
~\cite{Chabanier:2019eai,2016ApJ...824L..31B,2017PhRvL.119d1102K,Stott:2018opm,Unal:2020jiy,Kimball2023}. Furthermore, we demonstrate that the same ALP models that address the JWST excess may yield complementary signals in a variety of current/future axion search experiments ~\cite{Berlin:2020vrk,Graham:2020kai,Bloch:2019lcy}, and intriguingly have the potential to alleviate puzzling features found in the small-scale structure of DM halos~\cite{Tulin170502358,Kaplinghat:2015aga,Bullock:2017xww}. In addition, we identify a parameter range that is consistent with existing $\Lambda$CDM predictions, which is worth exploration even if the current JWST excess resolves upon further investigations. 

\section{Oscillatons from delayed ALP oscillation} 
\label{sec:delay}
Oscillatons are bound objects of a real scalar field (such as an ALP) sustained by self-interaction and gravity~\cite{Soda:2017dsu,Kitajima:2018zco,Olle:2019kbo,Brax:2020oye}. They are commonly referred to as oscillons when the effect of gravity is negligible~\cite{Urena-Lopez:2001zjo,Marsh:2015xka,Bogolyubsky:1976nx,Copeland:1995fq,Fodor:2006zs,2011PhRvD..84d3531C,Mukaida:2016hwd,Zhang:2018slz,Eby:2018ufi,Olle:2020qqy,Olle:2019kbo}. Since real scalar fields do not possess a conserved charge, the associated oscillons have a finite lifetime and can decay within the age of the universe. Nevertheless, an approximate $U(1)$ symmetry that appears in the non-relativistic regime makes them relatively long-lived~\cite{Mukaida:2016hwd,Eby:2018ufi,Zhang:2020bec,Olle:2020qqy,Visinelli:2021uve}. 
Once they survive into the matter-dominated era, the effect of gravity may further stabilize their configurations, enabling them to persist until the present day~\cite{Fodor:2009kg,Olle:2020qqy,Olle:2019kbo}. 

ALP is one of the most well-motivated examples of a real scalar field due to its connections to QCD axions and string theories. 
Axions are pseudo-Nambu-Goldstone bosons arising from spontaneous breaking of global Peccei-Quinn (PQ) symmetries, originally proposed to solve the strong CP problem in QCD ~\cite{Peccei:1977hh,Peccei:1977ur,Wilczek:1977pj}. Later on, they were found to be attractive dark matter candidates, and more general non-QCD ALPs are also well motivated from theoretical frameworks such as string theory and supersymmetry~\cite{Weinberg:1977ma,Abbott:1982af,Dine:1982ah,Preskill:1982cy}. The very light $m_a$ needed to affect high redshift structure growth as we found is incompatible with the QCD axion, but is generic for ALPs. 

Axions can be produced from both misalignment mechanisms (MM) and the decay of axion topological defects. The axion energy density resulting from the latter mechanism is contingent upon model specifics and still under development, with uncertainties in part due to technical challenges in simulations (see e.g.  \cite{LISACosmologyWorkingGroup:2022jok,Chang:2021afa,Chang:2019mza, Ellis:2022grh,Xiao:2021nkb,Eggemeier:2019khm,Buschmann:2019icd,Chang:2023rll}, and the refs therein). In particular, the effect of topological defects can be absent in certain cosmological scenarios. For example, if PQ symmetry breaking occurs before inflation, axion cosmic strings would be sufficiently diluted; on the other hand, the contribution from axion domain walls can be inconsequential when compared to that from the MM mechanism, if the lifetime of domain walls is sufficiently short. For these reasons, here we choose to focus on the formation of oscillatons via the misalignment mechanism.

In the standard misalignment mechanism (SMM), the axion field is initially displaced from its true potential minimum with zero initial velocity. It starts to oscillate when it acquires a mass $m_a$ similar to the Hubble rate, $H_*\sim m_a$. There is no oscillon formation, and the coherent oscillations of the axion field contribute to the relic abundance of cold dark matter (CDM) today ($\Omega_a$). 

In contrast, when oscillons are formed, they typically arise from a delayed onset of ALP oscillations with parametric resonances. 
Our analysis assumes a significant delay in the onset of these oscillations and the subsequent formation of long-lived oscillons that persist until the present time. 
However, the specific model details leading to this scenario do not change our results. 

Several mechanisms delaying the onset of ALP field oscillations have been investigated in the literature.
For example, the ALP field may initially slow-roll in a plateau potential, remaining homogeneous until it reaches a deeper potential at later times. Such plateau potentials have been investigated in models motivated by $\alpha$ attractors and axion monodromy~\cite{Soda:2017dsu,Kitajima:2018zco,Olle:2019kbo,Brax:2020oye,Olle:2020qqy}.
Alternatively, the initial field velocity is another approach to delay the onset of oscillation and has been explored in the context of kinetic misalignment (KMM)~\cite{Co:2019jts, Chang:2019tvx}. 
In these models, oscillons also arise from parametric resonances. We will elaborate on such models with simulations in a forthcoming paper~\cite{future}. 
Meanwhile, several other models following analogous spirits have been explored in the literature such as large misalignment~\cite{Arvanitaki:2019rax}, trapped misalignment~\cite{DiLuzio:2021gos,Nakagawa:2020zjr}, and KMM with radial excitations~\cite{Co:2021lkc,Co:2019jts,Co:2019wyp}. They all lead to some amount of delayed oscillations but do not span a large enough parameter space to enable all the DM energy density to go into oscillatons. 

In this study, we focus on model-independent aspects of the delayed onset of oscillations and elaborate on its implications for JWST observations. The generic features are largely determined by two physical parameters: the mass of the ALP particle $m_a$ and a dimensionless time parameter $\eta \equiv m_a/H_*$, which quantifies the degree of oscillation delay. A delayed oscillation corresponds to a larger value of $\eta$, compared to approximately $3$ in the SMM. The fraction of the field that fragments saturate quickly with increasing $\eta$. To ensure that the ALP field is fully fragmented and forms oscillons, we focus on the parameter space where $\eta>40$.

While $\eta$ and $f_a$ are in principle independent model parameters, for a given model with $m_a$ and $\Omega_a$ specified, they have a $1-1$ relation.
As an example, for a simple, temperature-independent axion potential as used in most ALP models, $f_a$ relates to $\eta$ in the following way~\cite{Eroncel:2022efc}:
\begin{eqnarray}
\label{eq:fa}
f_a\approx 10^{15}~{\rm GeV} \left(\frac{10^{-20} {\rm eV}}{m_a} \right)^{\frac{1}{4}} \left(\frac{90}{\eta} \right)^{\frac{3}{4}} \left(\frac{g_*}{4} \frac{\Omega_a h^2}{0.12} \right)^{\frac{1}{2}},
\end{eqnarray}
where $g_*$ is the effective degrees of freedom in the entropy evaluated at the onset of fragmentation. Throughout our paper, we will assume that all the DM is comprised of the oscillatons, i.e. $\Omega_a=\Omega_{\rm DM}$, to fix $\eta$, which then fixes the $1-1$ correspondence between $f_a$ and $\eta$ for a given $m_a$.

The main features of the oscillatons can be specified given the $m_a$ and $\eta$ parameters. 
Their mean separation can be determined by the comoving scale $d_{\rm osc} = \pi/k_{\rm osc}$, where $k_{\rm osc}\approx m_a a_*\kappa_p$, with $a_*$ being the scale factor at the onset of oscillation and $\kappa_p\sim O(1)$ a correction factor that we will specify later in Eq.~(\ref{eq:kp}) to incorporate a mild $\eta$ dependence. 
The typical mass of the oscillaton can be estimated as
\begin{eqnarray}
\label{eq:m0}
M_{0}&\approx& \bar{\rho}_a \frac{4\pi}{3} \left(\frac{\pi}{m_a a_* \kappa_p} \right)^3 \\ \nonumber
 &\approx& 10^4~\rm M_{\odot}  \left( \frac{70}{\eta} \right)^{\frac{3}{2}+0.66~\theta(\eta-80)} \left(\frac{4 \times 10^{-20}~{\rm eV}}{\it{m_a}} \right)^{\frac{3}{2}},  
\end{eqnarray}
where $\theta$ is the Heaviside Function, $\bar{\rho}_a$ is the average axion energy density and is taken to be the average DM energy density today.
Both $\eta$ and $m_a$ anti-correlate with $M_0$, i.e. $M_0$ decreases as $\eta$, $m_a$ increases.
For a fixed $M_0$, increasing $\eta$ can accommodate a smaller $m_a$, which can help alleviate the BHSR constraints.
These axion clusters are produced from the collapse of the axion fields that fragmented at subhorizon scales.

The substantial mass of oscillatons naturally induces significant velocity dependent gravitational scatterings~\cite{Loeb:2022adr}.
In terms of the oscillaton mass, we have
\begin{eqnarray}
\label{eq:xsg}
    \frac{\sigma}{M_0} \approx 10 {\ \rm cm^2/g} \left(\frac{M_0}{10^4 {\rm M_{\odot}}}\right) \left(\frac{10\ {\rm km/s}}{v}\right)^4,
\end{eqnarray}
where $v$ is the relative velocity between the two oscillatons and anomalies in DM small scale structure observations may be addressed by values of $O(1)~\rm cm^2/g$  ~\cite{Tulin170502358,Kaplinghat:2015aga,Bullock:2017xww,2011MNRAS.415L..40B,2012MNRAS.422.1203B,2012MNRAS.423.3740V,Rocha:2012jg,2013MNRAS.431L..20Z,Peter:2012jh,2010MNRAS.406.1220W,Kaplinghat:2019svz,PhysRevLett.125.111105,Yang:2021kdf,Turner201002924,Correa:2022dey,Yang221113768}.
The wave nature of ALPs in our considered mass range gives rise to pc to kpc scale solitonic cores in ALP halos, which may also help alleviate small scale challenges in $\Lambda$CDM ~\cite{Hu:2000ke,Marsh:2015xka,Niemeyer:2019aqm}.
Given a large cross section, e.g., $10~\rm cm^2/g$, close encounters between oscillatons can merge them into larger, more diffuse structures within the ALP halos, enriching the small scale features of our model. 
The diffuse nature of the oscillatons also alleviates
the existing stringent constraints from stellar dynamical, microlensing, and small-scale structures that apply to massive compact halo objects such as primordial black holes~\cite{2016ApJ...824L..31B,2017PhRvL.119d1102K,Macho:2000nvd,EROS-2:2006ryy,2011MNRAS.416.2949W,Blaineau:2022nhy,Esteban:2023xpk}.

\begin{figure}[t]
\includegraphics[width=0.465\textwidth]{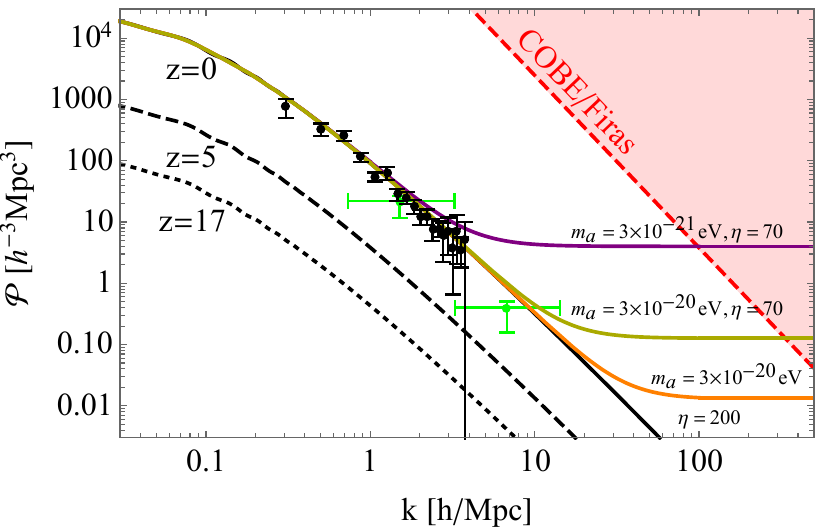}
\caption{\label{Fig1: Spectrum}
Example MPSs in delayed ALP oscillation models (colored curves) compared to the MPS from a standard $\Lambda$CDM model. Data points with error bars represent Lyman-$\alpha$ measurements (black) from Ref.\cite{Chabanier:2019eai} and HST UV luminosity function measurements (green) from Ref.\cite{Sabti:2021unj}. The red dashed curve denotes the maximal cut-off scale for $k^4$ growth in the adiabatic curvature power spectrum based on the COBE/FIRAS bound from Ref.\cite{Chluba:2013dna}. Higher resolution Lyman-$\alpha$ surveys can extend the scales on which the power spectrum is measured to $k\sim 10$ h/Mpc, similar to the UV luminosity data \cite{Irsic:2019iff,Rogers:2020ltq}.
}
\end{figure}

\section{Matter power spectrum}
We decomposed the density perturbation into linear and quadratic components following the methodology outlined in Ref.~\cite{Eroncel:2022efc}. 
The linear component approximately corresponds to the adiabatic perturbation in $\Lambda$CDM cosmology, denoted as $\mathcal{P}_{\rm CDM}(k)$. On the other hand, the quadratic component introduces a shot noise-like contribution referred to as $\mathcal{P}_{\rm osc}(k)$ which can be estimated as~\cite{Liu:2022bvr,Inman:2019wvr,1974A&A....37..225M,Dike:2022heo}
\begin{eqnarray}\label{eq:PAC_est}
{\cal P}_{\rm osc}(k) \approx (D_{\rm iso}(0)-1)^2 \frac{f_{\rm osc} M_0}{\rho_c \Omega_a} \propto M_0 \text{,~ for } k< k_{\rm osc}.\\ \nonumber
\end{eqnarray}
 Here $\rho_c$ is the critical energy density today, $f_{\rm osc}$ is the fraction of dark matter comprised by oscillatons which we fix to be $1$, $f_{\rm osc}\rho_c \Omega_a/M_0$ is the number density of the oscillatons, and 
 $D_{\rm iso}(z)$ is the growth factor of isocurvature perturbations which can be parameterized as~\cite{1974A&A....37..225M,Inman:2019wvr,Dike:2022heo}
\begin{eqnarray}
\label{eq:Diso}
D_{\rm iso}(z) &=& \left(1+\frac{\Omega_a}{\Omega_m a_-} \frac{3 a}{2 a_{\rm eq}} \right)^{a_-}, 
\end{eqnarray}
where 
\begin{eqnarray}
a_-&=&(\sqrt{1+24\Omega_a/\Omega_m}-1)/4.
\end{eqnarray}
The $D_{\rm iso}$ does not evolve at $a\ll a_{\rm eq}$ and grow as $3a/(2 a_{\rm eq})$ (identical to the adiabatic one) for $\Omega_a=\Omega_m$ at $a\gg a_{\rm eq}$. 
If oscillatons comprise only a fraction of dark matter, then $f_{\rm osc} < 1$, resulting in a smaller ${\cal P}_{\rm osc}(k)$. 
Note that the oscillaton number density is independent of $\Omega_a$ because $M_0\propto \Omega_a$. However, ${\cal P}_{\rm osc}$ decreases as $\Omega_a$ decreases through the $D_{\rm iso}$ in Eq. \ref{eq:Diso} and Eq.~\ref{eq:PAC_est}, resulting in fewer high-z massive galaxies. 

We numerically calculate the leading order ${\cal P}_{\rm osc}(k)$, including backreaction, following Eq.~$4.11$ in Ref.~\cite{Eroncel:2022efc}, and parameterize the obtained results as \footnote{
We solved for the relic abundance neglecting the effect of fragmentation which could induce a sub-{\cal O}(1) uncertainty~\cite{Eroncel:2022vjg}.
A more accurate determination of the relic abundance requires dedicated study which is beyond the scope of this work. 
}:
\begin{align}
\label{Eq: Axion Spectrum}
    \mathcal{P}_{\rm osc}(k) = \frac{2 \pi^2}{k^3} \frac{ f(\eta) (D_{\rm iso}(0)-1)^2}{(D_{\rm iso}(z_{\rm eq})-1)^2} \left( \frac{\kappa}{\kappa_p(\eta)} \right)^3 \theta(\kappa_p - \kappa), 
\end{align}
where $\kappa \equiv \frac{k}{m_a a_*}$,  $\kappa_p(\eta)$ and $f(\eta)$ are numerical factors parameterized as
\begin{align}
\label{eq:kp}
    \kappa_p(\eta) \simeq \left\{            
\begin{aligned}
& 0.9, \;\;\;\;\;\;\;\;\;\;\;\;\;\;\;\;\;\;\;\;\;\;\;\;\;\;\;\;\;\;\;\;\;\; 40 \leq \eta \leq 80, \\
&0.9\left( \frac{\eta}{80} \right)^{0.22}, \;\;\;\;\;\;\;\;\;\;\;\;\;\;\;\;\;\;\, 80 <\eta \leq 10^3 , 
\end{aligned}
\right.
\end{align}
and
\begin{align}
    f(\eta) \simeq  1.19\left( \frac{\eta}{80}\right)^{-0.1}, \;\;\;\;\;\;\;\;\;\;\;\;\;\;\;\;\;\;\, 40 <\eta \leq 10^3. 
\end{align}
The spectrum $\mathcal{P}_{\rm osc}(k)$ becomes dominant at scales larger than $k_*\approx a_* H_*$ and is truncated at $k_{\rm osc}$. 
For $k > k_{\rm osc}$, the spectrum becomes negligibly small and decreases as $1/k^2$ \cite{Eroncel:2022efc,Eroncel:2022vjg}.

In Fig.~\ref{Fig1: Spectrum}, we show the MPS for three benchmark examples. 
The $M_0$ is $1.6\times 10^3$, $1.5\times 10^4~\rm M_{\odot}$, and $4.9\times 10^5$ for the orange, golden, and magenta cases, respectively. 
We will show in Fig.~\ref{Fig3: 1sigma} that the case in gold color is favored in order to explain the JWST excess, the orange case is consistent with $\Lambda$CDM, while the magenta case overproduces structures and is excluded by measurements of the UV luminosity function. 
In all cases, ${\cal P}_{\rm osc}(k) \gg {\cal P}_{\rm CDM}(k)$ for $k > k_*$, where the spectrum plateaus, before being truncated at $k \sim k_{\rm osc}$ (beyond the range of Fig.~\ref{Fig1: Spectrum}). 
The case in gold color, specifically, has a transition around $k_*\sim 10~{\rm h/Mpc}$, which is close to the scale associated with the JWST excess. 

\begin{figure}[tb]
\includegraphics[width=0.495\textwidth]{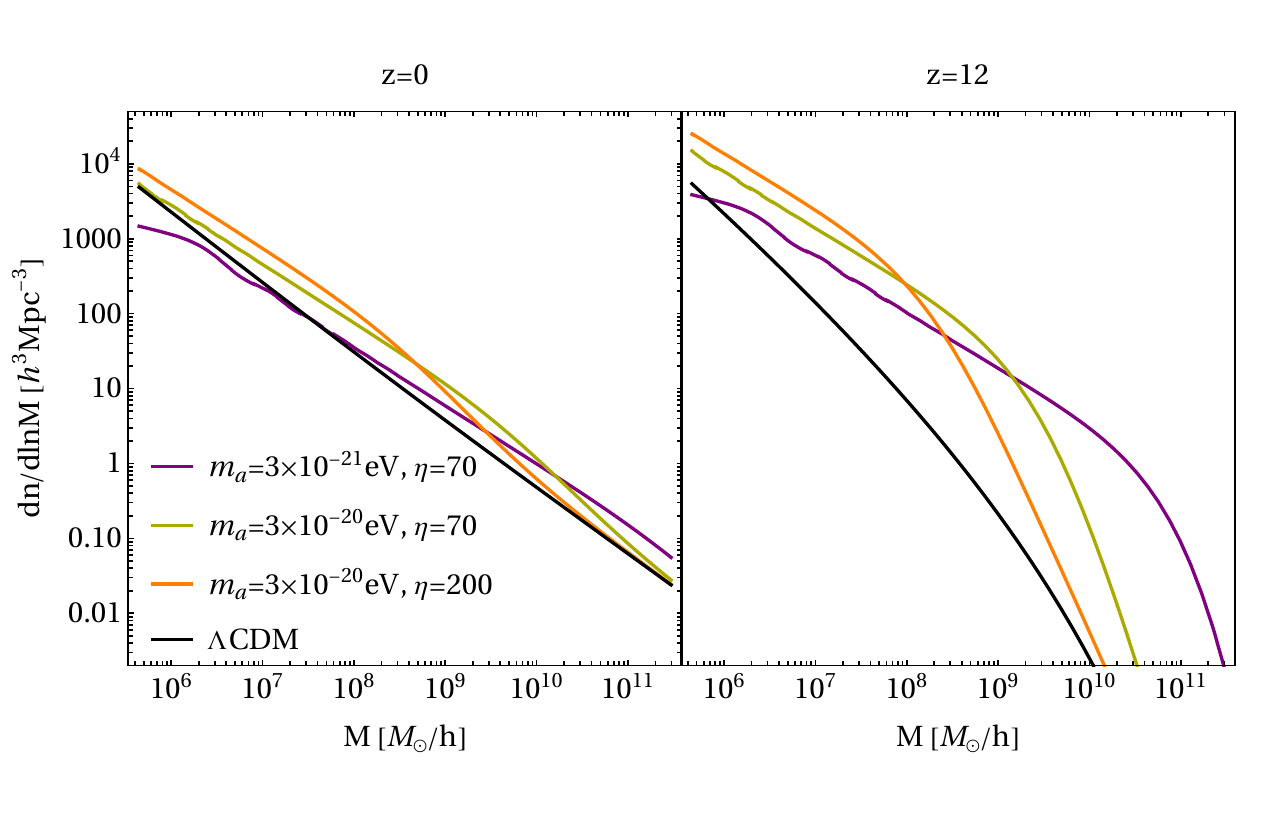} 
\caption{\label{Fig2: Formation} The halo mass function for the benchmark cases of Fig.~\ref{Fig1: Spectrum} at redshifts $z=0$ and $z=12$. }
\end{figure}

\section{Halo Mass Function} 
The clustering of oscillatons in the matter era leads to the formation of halos. We use the Press-Schechter formalism to estimate the halo mass function (HMF) from the MPS~\cite{1974ApJ...187..425P}.
Given an MPS ${\cal P}(k)$ at redshift zero, we compute the mass variance using a top-hat window function as
$$
\sigma^2_M(R) = \dfrac{1}{2\pi^2} \int d k {\cal P}(k) \left( 3 \dfrac{\sin(k R)-k R \cos(k R) }{k R} \right)^2 k^2, 
$$
where $M$ and $R$ are halo mass and radius and are related through $M=4\pi\bar{\rho}_m R^3/3$, with $\rho_m$ being the mean matter density today, 
and the redshift dependence of $\sigma_M$ is conventionally absorbed into the definition of $\delta_c(z)$. 
The HMF is computed in terms of $\nu\equiv \delta_c^2(z)/\sigma^2_M$ as
\begin{eqnarray}
\label{eq:hmf}
\frac{d n}{d M} = \frac{\bar{\rho}_m}{M} \nu f_{\rm EPS}(\nu) \frac{d\ln \nu}{d\ln M},    
\end{eqnarray}
where $\nu f_{\rm EPS}(\nu)$ is a shape function including the effect of ellipsoidal collapse from Ref.~\cite{Sheth:1999mn}.
This approach has been widely applied in the literature because of its simplicity.
Ref.~\cite{Liu:2022okz} demonstrated that this technique can produce reliable approximations of the N-body reconstructed HMFs, even in the presence of massive primordial black holes.

In Fig.~\ref{Fig2: Formation}, we show the obtained HMFs for the $\Lambda$CDM and the three benchmarks in Fig.~\ref{Fig1: Spectrum} at $z=0$ (left) and $z=12$ (right). We see that the HMFs at $z=0$ are relatively similar across all cases; however, at $z=12$, the delayed oscillation cases exhibit significant enhancements towards larger halo masses, increasing the likelihood of finding halos hosting the very massive galaxies.
Once small halos virialize, they almost decouple from the background evolution. Hence the abundance of massive halos is largely unaffected at low redshifts.
As a result, the effect of our model becomes less significant today, which alleviates constraints from low redshift measurements~\cite{Dike:2022heo,Esteban:2023xpk}.


\begin{figure}[t]
\includegraphics[width=0.48\textwidth]{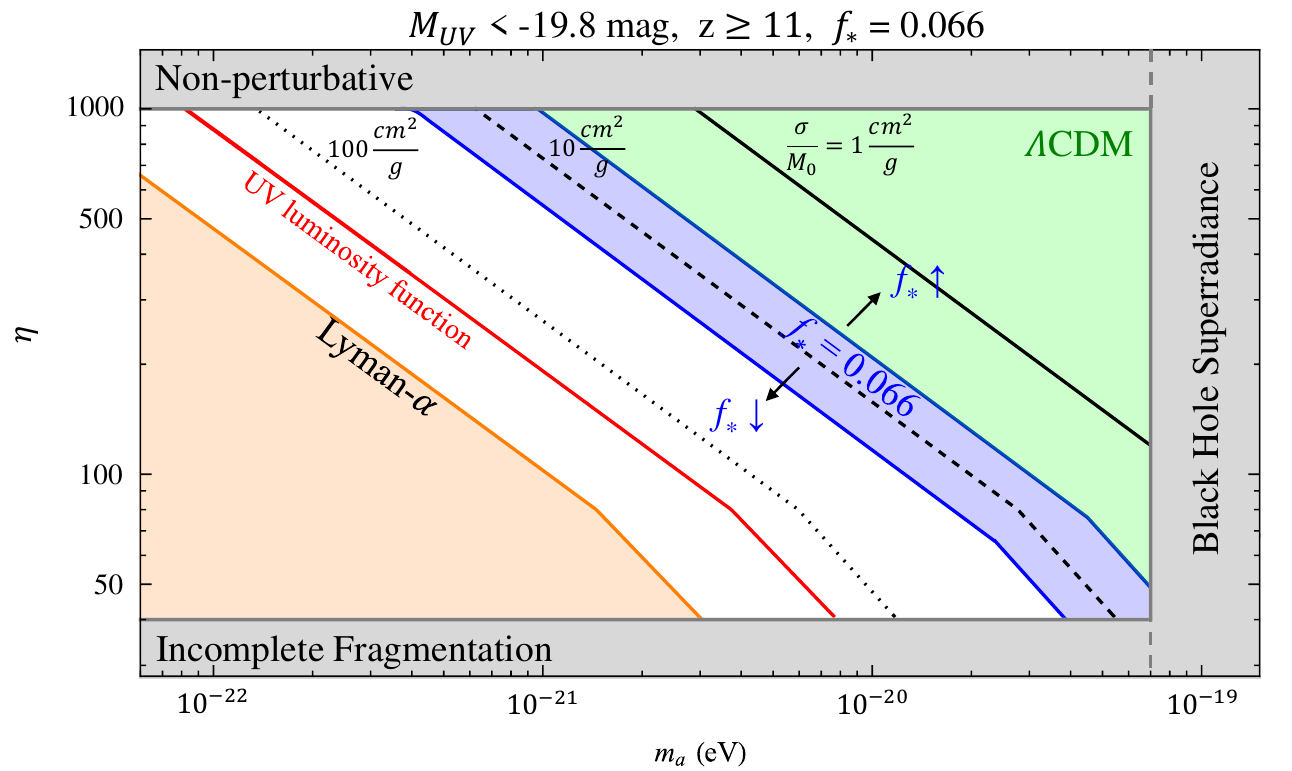} 
\caption{\label{Fig3: 1sigma} The viable model parameter space given the observations in Ref. \cite{2023ApJS..265....5H}, considering galaxies with $M_{\rm UV}<-19.8~\rm mag$ at $z \geq 11$. 
The 68\% CL preferred region shaded in blue corresponds to a prediction of $0.74<N_{\rm exp}<4.3$. It is obtained for a star formation efficiency $f_* = 0.066$ and would shift up (down) if $f_*$ increases (decreases). 
Black lines represent contours of the gravitational scattering cross section per mass for values of $1,~10,$ and $100\rm cm^2/g$, evaluated at a velocity of $v=10~\rm km/s$. 
At larger velocities, the cross section quickly reduces due to the $v^{-4}$ dependence. 
Constraints from various sources exclude certain regions: the Lyman-$\alpha$ forest data \cite{Chabanier:2019eai} disfavors the orange area; the region to the left of the red curve is inconsistent with measured UV luminosity functions \cite{Sabti:2021unj}; and the grey area to the right is excluded by BHSR constraints \cite{Arvanitaki:2014wva, Stott:2018opm,Unal:2020jiy,Kimball2023}.
}
\end{figure}

\begin{figure*}[t]
\includegraphics[width=0.49\textwidth]{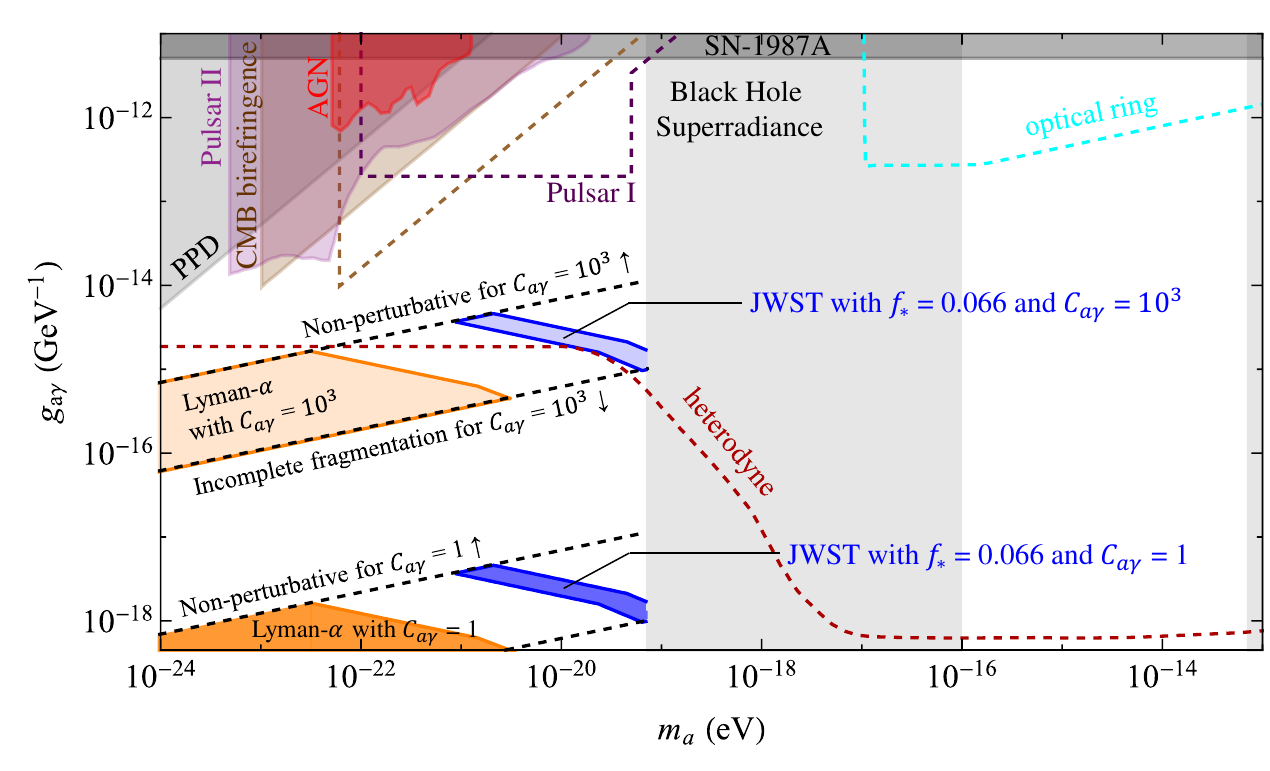} 
\includegraphics[width=0.49\textwidth]{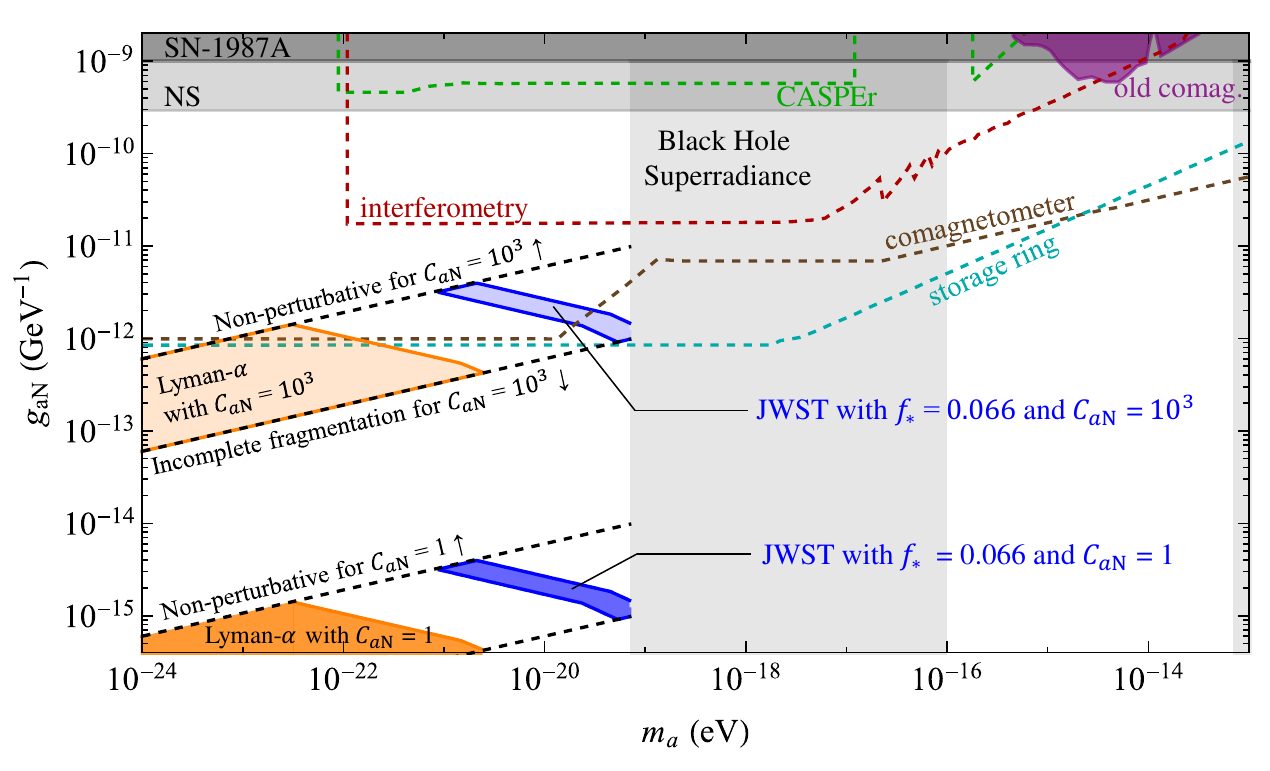} 
\caption{\label{Fig4} Axion-photon (left) and axion-nucleon (right) couplings versus axion DM mass $m_a$ (assuming $\Omega_a=\Omega_{\rm DM}$ in Eq.~\ref{eq:fa}). The blue and the lighter-blue regions correspond to the regions favoured by matching the JWST excess, with $C_{a\gamma,N} = 1$ and $C_{a\gamma,N} = 10^3$, respectively. The blue regions are bounded by two dashed black lines, corresponding to the requirement of perturbative $\eta$ with complete fragmentation. Constraints from various existing searches (solid shaded regions) and forecasts (colored dashed lines) are also illustrated. See text for further details.}
\end{figure*}

\section{Parameter space favored by JWST excess}
The enhanced massive halo population as we see in Fig.~\ref{Fig2: Formation} can trigger earlier galaxy formation and enhance the formation efficiency~\cite{Zurek:2006sy,Dekel:2023ddd}. 
To convert the model prediction for HMF into the number of observed galaxies, we consider a simplified approach assuming a constant star formation efficiency $f_*$. 
We compute the expected number of observed galaxies with $M_{\rm UV}<-19.8~\rm mag$ at $z> 11$ as considered in Fig. 13 of Ref. \cite{2023arXiv230406658H}, where the number of observed galaxies $N_{\rm obs}=2$ is systematically higher than theoretical predictions from the literature~\cite{2014MNRAS.445.2545D,2019MNRAS.486.2336D,2020MNRAS.496.4574Y,2020MNRAS.499.5702B,2023MNRAS.519.3118W,2023MNRAS.521..497M}, e.g., $0.3$ from Ref.~\cite{2023arXiv230404348Y} as a recent estimate. We convert the requirement of $M_{\rm UV}<-19.8~\rm mag$ into a minimum halo mass by using the $M_*-M_{\rm UV}$ relation of Ref.~\cite{2023arXiv230404348Y}. 
Assuming a constant star formation efficiency $f_*$, we set the minimum halo mass $M_{\rm min} = 1.8 \times 10^9 e^{-0.12 z_{\rm min}}/(f_* f_b)~\rm M_{\odot}$, where $z_{\rm min}=11$, $f_b\approx 0.16$ is the baryon fraction in the matter energy density, and we set $f_*=0.066$ so that the $\Lambda$CDM prediction is for an expected $0.3$ galaxies at $z > 11$~\cite{2023arXiv230404348Y}. 
The value of $f_*=0.066$ is within expectations for a $10^{12}~\rm M_{\odot}$ halo at z=0, but small compared to some theoretical models proposed to explain the JWST excess solely by extra star formation~\cite{Dekel:2023ddd,Boylan-Kolchin:2022kae,2023MNRAS.518.2511L,2023MNRAS.521..497M,2023MNRAS.519..843M}.
It has been noted that an excessively high $f_*$ could introduce new tensions with the cosmic reionization history~\cite{Gong:2022qjx}, making our obtained value a more desirable option. 
The number of expected galaxies count is calculated as  
\begin{eqnarray}
    N_{\rm exp} = \frac{A_{\rm eff}}{4\pi} \int_{11}^{z_{\rm max}} d z \frac{d V_c}{d z} \int_{M_{\rm min}}^{M_{\rm max}} d M \frac{dn}{dM},
\end{eqnarray}
where $A_{\rm eff}=45 ~{\rm arcmin^2}$ is the effective area of the observation following Ref. \cite{2023arXiv230406658H} and $V_c$ is the comoving Hubble volume at $z$. $dn/dM$ is the HMF in Eq.~\ref{eq:hmf}.
Our result is not sensitive to $z_{\rm max}$, $M_{\rm max}$, provided that they are sufficiently large.

Fig. \ref{Fig3: 1sigma} presents the preferred $68\%$ CL parameter regions for a constant $f_*=0.066$ as a blue band in the $m_a-\eta$ plane. We obtain the blue band considering $0.74<N_{\rm exp}<4.3$ which arises from a Feldman-Cousins analysis for two observations and zero background~\cite{Feldman:1997qc}, and the lower limit happens to coincide with that from Ref. \cite{2023arXiv230406658H}. 
For a higher (lower) value of $f_*$, the band would shift upward (downward). 
The upper-right corner corresponds to lower prediction for $N_{\rm exp}$ values that are consistent with $\Lambda$CDM  predictions within the uncertainties~\cite{2014MNRAS.445.2545D,2019MNRAS.486.2336D,2020MNRAS.496.4574Y,2020MNRAS.499.5702B,2023MNRAS.519.3118W,2023MNRAS.521..497M}. 
The lower-left region has higher $N_{\rm exp}$ and is constrained by Lyman-$\alpha$ and UV luminosity measurements~\cite{Chabanier:2019eai,Sabti:2021unj}. 
We also plot the contours of the gravitational scattering cross section of Eq. (\ref{eq:xsg}) evaluated at $v=10~\rm km/s$ in Fig. \ref{Fig3: 1sigma}. 
In the viable region, we see that the gravitational cross section takes values in $(1-100)~\rm cm^2/g$, which could lead to novel signal predictions that can address the small-scale observations, e.g., the core-cusp problem and the too-big-to-fail problem~\cite{Tulin170502358,Kaplinghat:2015aga,Bullock:2017xww,2011MNRAS.415L..40B,2012MNRAS.422.1203B,2012MNRAS.423.3740V,Rocha:2012jg,2013MNRAS.431L..20Z,Peter:2012jh,2010MNRAS.406.1220W,Kaplinghat:2019svz,PhysRevLett.125.111105,Yang:2021kdf,Turner201002924,Correa:2022dey,Yang221113768}.

\section{Implications for Axion Models and Complementary Searches} 
As shown in the last section, axion models with delayed ALP oscillations can address the JWST excess, with a preferred parameter space spanning $4\times 10^{-22}~{\rm eV}<m_a<10^{-19}~\rm eV$ and $40<\eta<1000$. Existing ALP searches may provide valuable complementary probes for such ALPs, albeit in a model-dependent way. While the impact of ALP on structure formation is largely independent of the non-gravitational axion coupling to the Standard Model (SM) particles, most other ALP searches do depend on the specific coupling patterns. Here we focus on axion interactions with photons and nucleons ($N$), as described by
\begin{eqnarray}
\mathcal{L} \,\in\, \frac{g_{a\gamma}}{4} a F_{\mu \nu} \tilde{F}^{\mu \nu} + g_{aN} \partial_{\mu} a \bar{N}\gamma^{\mu}\gamma^5 N,
\end{eqnarray}
with effective couplings
\begin{align} \label{eq:Cs}
    g_{a\gamma} = C_{a\gamma} \alpha/(2\pi f_a), \;\;\;\;\;\;\;\;\;\;
    g_{aN}=C_{aN}/f_a,
\end{align}
where $F_{\mu \nu}$ represents the electromagnetic field strength, and $C_{a\gamma}$, $C_{aN}$ are model-dependent parameters. In standard QCD axion inspired models such as KSVZ or DFSZ \cite{Kim:1979if,Shifman:1979if,Dine:1981rt,Zhitnitsky:1980tq}, 
$C_{a\gamma}$, $C_{aN}\sim O(1)$. Meanwhile, other recently proposed well-motivated axion theories such as those with multiple PQ fermions, vector kinetic mixing, and axion clockworks, can generically predict significantly larger coupling $C$'s, up to $\mathcal{O}(10^{2} - 10^3)$ \cite{Dror:2020zru, Agrawal:2017cmd, Agrawal:2018mkd}. 
According to Eqs.~\ref{eq:fa} and \ref{eq:Cs}, with our assumption of $\Omega_a=\Omega_{\rm DM}$, $f_a$ and the corresponding $\eta$ are fixed for given $g_{a\gamma}~(g_{aN})$ and $C_{a\gamma}~(C_{aN})$. Thus we may map the parameter regions in Fig.~\ref{Fig3: 1sigma} onto the $m_a-g_{a\gamma, N}$ space as in Fig.~\ref{Fig4}. In light of the aforementioned theoretically motivated range for $C_{a\gamma},~C_{aN}$,
in Fig.~\ref{Fig4}, the JWST excess favored regions are illustrated for two benchmark values: $C_{a\gamma},~C_{aN}= 1$ (blue band) and $C_{a\gamma},~C_{aN}=1000$ (light blue band). The viable regions for other values of $1\lesssim C_{a\gamma},~C_{aN}\lesssim 1000$ are expected to lie between these two bands.  These bands are truncated from above and below (denoted by dashed black lines) by requiring that the corresponding $\eta$ are within the perturbative regime with complete fragmentation, $40\lesssim\eta\lesssim1000$ (see discussion above Eq.~\ref{eq:fa}), which we have chosen to focus on in our analysis. As $C_{a\gamma}$ or $C_{aN}$ increases, the blue and orange bands, along with the black dashed lines, would linearly shift upwards. We also show the current constraints (solid shaded regions) and future sensitivity forecast (bounded above by colored dashed lines) for a variety of experiments/observations, including Lyman-$\alpha$~\cite{Chabanier:2019eai},  BHSR~\cite{Stott:2018opm,Unal:2020jiy,Kimball2023}, Supernova-1987A~\cite{Payez:2014xsa}, neutron star cooling (NS)~\cite{Chang:2018rso,Beznogov:2018fda,Hamaguchi:2018oqw,Sedrakian:2015krq}, protoplanetary disk polarimetry (PPD)~\cite{Fujita:2018zaj}, active galactic nuclei (AGN)~\cite{Ivanov:2018byi}, and old comagnetometer~\cite{Bloch:2019lcy}. 

As can be seen from Fig.~\ref{Fig4}, upcoming experiments can offer complementary probes for the parameter range that could address the JWST excess. For example, for the benchmark of $C_{a\gamma}=1000$ the reach of future heterodyne experiments~\cite{Berlin:2020vrk} overlaps with the JWST excess favored region of ($g_{a\gamma}$, $m_a$). Similarly, future storage ring and comagnetometer experiments would be able to explore the ($g_{aN}$, $m_a$) range motivated by the JWST excess~\cite{Graham:2020kai,Bloch:2019lcy}. It is also worth noting that the experiments covering the further upper regions in Fig.~\ref{Fig4}, such as interferometry \cite{Graham:2017ivz} and linearly polarized pulsar light (Pulsar I in Fig.\ref{Fig4} left) \cite{Liu:2019brz}, can potentially probe ALPs with $\eta>10^3$ and $C_{a\gamma}$ (or $C_{aN}$) $>10^3$. However, these regions require a non-perturbative analysis of fragmentation ($\eta \gtrsim 10^3$), which is beyond the scope of this study. \\

We note that in our preferred parameter range, the onset of delayed axion oscillation occurs at $T_*\lesssim 10~\rm keV$. Such a late onset is potentially subject to constraint from Big Bang Nucleosynthesis (BBN) \cite{Eroncel:2022vjg} if the dark sector induces a sufficiently large change in the energy density. For most mechanisms of realizing a delayed oscillation as outlined in Section~\ref{sec:delay}, the effect on BBN is negligible, such as the $\alpha$ attractor and axion monodromy models that implement the delay through a long flat plateau region in the potential~\cite{Soda:2017dsu,Kitajima:2018zco,Olle:2019kbo,Brax:2020oye}.
A potentially notable impact on BBN would only occur in minimal KMM models without regulating the extrapolation of kinetic energy towards arbitrarily higher temperatures or earlier times~\cite{Eroncel:2022vjg}. However, in realistic models, the dominance of axion kinetic energy must be transient, initially preceded by the dominance of another form of energy, which may provide a natural cutoff to the kinetic energy domination and thus alleviate the BBN constraint. The results of this work is independent of such model specifics. We will elaborate related discussions based on specific models in a follow-up work~\cite{future}.
In synergy with our study, recently there has been a general rising interest in viable post-BBN dark sector dynamics, which can lead to observable features in supermassive black holes, primordial gravitational waves, and small-scale structures, offering potential insights into recent astrophysical observations~\cite{Oikonomou:2023qfz,Oikonomou:2023bah,Bringmann:2023opz,Gouttenoire:2023bqy,Gouttenoire:2023nzr,Ghosh:2023aum,Chakraborty:2022mwu,DiLuzio:2021gos,Hutsi:2022fzw,Nakagawa:2020zjr,Sarkar:2014bca,Brouzakis:2005cj}.

\section{ Conclusion} 
We demonstrated that there is a large parameter space of axion dark matter with
$10^{-22}\rm eV\lesssim {\it{m_a}}\lesssim 10^{-19}\rm eV$ that can address the JWST excess while being compatible with all existing constraints, in the framework where the axions undergo a delayed oscillation.
The delayed onset of the axion field oscillation allows for efficient axion field fragmentation at subhorizon scales. 
The fragmented axion field collapses into a large population of massive oscillatons, which leads to more massive galaxies at high redshift, and thus can potentially address the excess observed by JWST. 
Upcoming experiments can provide complementary probes for the ALP parameter range favored by the JWST excess. 
Near-future Lyman-$\alpha$ forest surveys such as Weave-QSO \cite{WeaveQSO} or DESI \cite{DESILya1, DESILya2} will extend the scales on which the MPS can be measured by a further factor of $2-3$. Future surveys of strong lensing caustics could directly detect the predicted signature from small halos \cite{2023arXiv230612864D}.
We also identified sizable gravitational scattering in our model parameter space, which enriches small scale structure formation in our model that is worth further investigation~\cite{Kaplinghat:2015aga,Bullock:2017xww}. 
In addition to the complementary astrophysical probes related to structure formation, we demonstrated that the JWST excess favored parameter region can be probed by existing or planned axion search avenues, e.g. heterodyne, comagnetometer and storage ring experiments, assuming certain patterns of ALP-SM couplings (i.e. fixed $C_{a\gamma}, ~ C_{aN}$). This study identifies new avenues for probing axion DM, and would stand as a worthwhile addition to the literature even if the current JWST excess resolves after further investigation.

\section{Acknowledgment}
We thank Yuichi Harikane, Huangyu Xiao, and Hai-Bo Yu for the helpful discussion. SB is supported by NASA ATP-80NSSC21K1840. CC and YC are supported by the US Department of Energy under award number DE-SC0008541. DY is supported by the John Templeton Foundation under grant ID \#61884 and the U.S. Department of Energy under grant No.\ de-sc0008541. The opinions expressed in this publication are those of the authors and do not necessarily reflect the views of the John Templeton Foundation.

\bibliographystyle{elsarticle-num}
\bibliography{References}

\end{document}